\documentclass{webofc}
\usepackage[varg]{txfonts}   
\usepackage{hyperref}
\usepackage{url}
\hypersetup{colorlinks=true,citecolor=blue,urlcolor=blue,linkcolor=blue}
\newcommand{\ba}{\begin{eqnarray}}
\newcommand{\ea}{\end{eqnarray}}

\begin{document}
\title{Algebraic treatment of $\alpha$-cluster nuclei}
%
%


\author{\firstname{Roelof} \lastname{Bijker}\inst{1}
\fnsep\thanks{\email{bijker@nucleares.unam.mx}}
\and \firstname{Omar Alejandro} \lastname{D{\'{\i}}az Caballero}\inst{1}\fnsep\thanks{\email{odiazcab@gmail.com}}}
        
\institute{Instituto de Ciencias Nucleares, 
Universidad Nacional Aut\'onoma de M\'exico, 
A.P. 70-543, 04510 Ciudad de M\'exico, M\'exico}

\abstract{In this contribution we review the algebraic cluster model (ACM) for $\alpha$-cluster nuclei with $A=4k$ nucleons and its extension to the cluster shell model (CSM) for $A=4k+x$ nuclei. Particular attention is paid to the question to what extent the $\alpha$-cluster structure survives under the addition of extra nucleons. As an example, we discuss the properties of $^{12}$C and $^{13}$C.}

\maketitle

\section{Introduction}

Cluster degrees of freedom are very important for the description of light nuclei due to the large binding energy of the $^4$He nucleus. Early work on $\alpha$-cluster models goes back to the 1930's with studies by Wheeler \cite{wheeler}, and Hafstad and Teller \cite{Teller}, followed by later work by Brink \cite{Brink1,Brink2} and Robson \cite{Robson1,Robson2}. Recently, there has been a lot of renewed interest in the structure of $\alpha$-cluster nuclei, especially for the nucleus $^{12}$C \cite{FreerFynbo}. The measurement of new rotational excitations of the ground state \cite{Freer2007,Kirsebom,Marin} and of the Hoyle state \cite{Itoh,Freer2012,Gai,Freer2011} has stimulated a large theoretical effort to understand the structure of $^{12}$C (for a review see {\it e.g.} Refs.~\cite{FreerFynbo,Schuck,Freer,ppnp}). 

In this contribution we review the algebraic cluster model (ACM) for $\alpha$-cluster nuclei with $A=4k$ nucleons and its extension to the cluster shell model (CSM) for $A=4k+x$ nuclei. The ACM and CSM provide a theoretical approach for the study of $\alpha$-clustering in light nuclei based on the discrete symmetry of the underlying geometric configuration of $\alpha$ particles. The ACM was introduced in a study of $^{12}$C in terms of a cluster of three $\alpha$-particles in a triangular configuration with ${\cal D}_{3h}$ symmetry \cite{acm}. The corresponding ground state rotational band is characterized by the sequence $0^{+}$, $2^{+}$, $3^{-}$, $4^{\pm}$ and $5^{-}$, including the prediction of a  parity doublet, $4^{+}$ and $4^{-}$, and the existence of a $5^-$ state which was observed experimentally more than a decade later \cite{Marin}. An interesting question is to what extent the $\alpha$-cluster structure survives under the addition of extra nucleons. These nuclei are discussed in the framework of the Cluster Shell Model (CSM) which describes the behavior of single-particles levels in a deformed field generated by the cluster of $\alpha$-particles \cite{csm}. As an example, we discuss the properties of $^{12}$C and the neighboring nucleus $^{13}$C \cite{prl122,epjst,plb843}.

\section{Algebraic Cluster Model}

The Algebraic Cluster Model (ACM) describes the relative motion of $n$-body clusters in terms of a spectrum generating algebra (SGA) of $U(\nu+1)$ where $\nu=3(n-1)$ represents the number of relative spatial degrees of freedom \cite{acm}. For the two-body problem the ACM reduces to the $U(4)$ vibron model \cite{cpl}, for three-body clusters to the $U(7)$ model \cite{acm,BIL} and for four-body clusters to the $U(10)$ model \cite{O16,RB}. The ACM has a very rich symmetry structure. In addition to continuous symmetries like the angular momentum, in case of $\alpha$-cluster nuclei the Hamiltonian has to be invariant under the permuation of the $n$ identical $\alpha$ particles. Since one does not consider the excitations of the $\alpha$ particles themselves, the allowed cluster states have to be symmetric under the permutation group $S_n$. The potential energy surface corresponding to the $S_n$ invariant ACM Hamiltonian gives rise to an equilibrium shapes of $\alpha$ particles located at the vertices of an equilateral triangle for $^{12}$C ($n=3$) and of a regular tetrahedron for $^{16}$O ($n=4$). In this contribution, we limit ourselves to the case of three $\alpha$-particles. 

\begin{figure}[t]
\centering
\begin{minipage}{.4\linewidth}
\includegraphics[width=\linewidth]{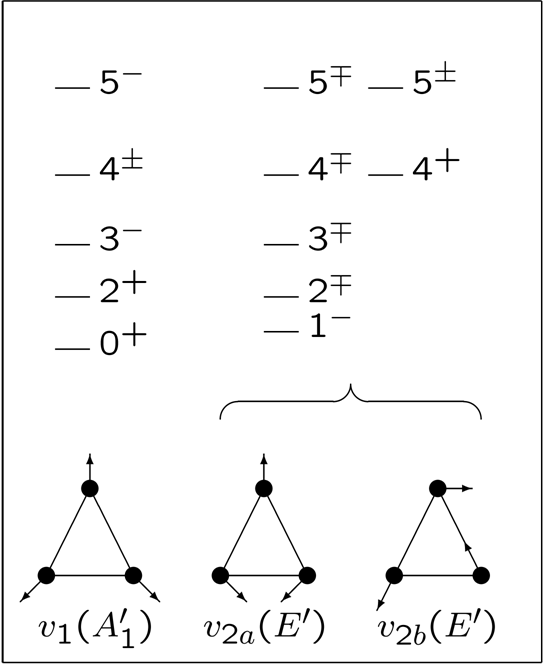}
\caption{Rotational bands with triangular symmetry}
\label{bands_d3h1}
\end{minipage} \hfill
\begin{minipage}{.55\linewidth}
\includegraphics[width=\linewidth]{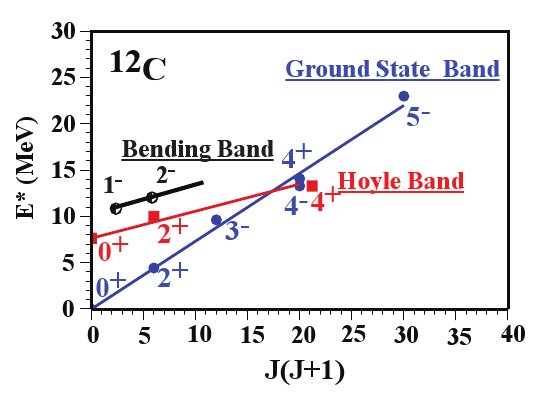}
\caption{Rotational bands in $^{12}$C. Taken from Ref.~\cite{Marin}.}
\label{bands_c12}
\end{minipage}
\end{figure}

The rotation-vibration spectrum of three $\alpha$-particles in a equilateral triangular configuration is to a good approximation given by 
\ba
E \;=\; \epsilon_{1}(v_{1}+\tfrac{1}{2}) + \epsilon_{2}(v_{2}+1) + \kappa \, L(L+1) ~.
\label{energy} 
\ea
There are two fundamental vibrations: a one-dimensional symmetric (or breathing) vibration $v_1(A'_1)$ and a two-dimensional vibration $v_2(E')$ consisting of a degenerate antisymmetric stretching vibration $v_{2a}$ and a bending vibration $v_{2b}$. The labels $A'_1$ and $E'$ refer to the transformation character under the corresponding point group ${\cal D}_{3h}$. Fig.~\ref{bands_d3h1} shows the structure of the rotational excitations: positive and negative parity states merge into a single rotational band. For example, the ground-state band with $(v_1,v_2)=(0,0)$ consists of the sequence $L^P=0^{+}$, $2^{+}$, $3^{-}$, $4^{\pm}$, $5^{-}$, $\ldots$, all of which have been observed in $^{12}$C, (blue circles in Fig.~\ref{bands_c12}). The Hoyle state and its rotational excitations are assigned to the breathing vibration $(v_1,v_2)=(1,0)$ (red) and the first $1^-$ state as the bandhead of the $(v_1,v_2)=(0,1)$ vibration (black). 

The equilateral triangular shape was confirmed in nuclear lattice effective field theory (NLEFT) calculations of the $0_1^{+}$, $2_1^{+}$, $3_1^{-}$, $4_1^{+}$ and $4_1^{-}$ states \cite{bonn,nature}. Similar calculations of the $0_2^{+}$ (Hoyle), $2_2^{+}$ and $4_2^{+}$ states showed a bent-arm configuration. If the Hoyle band would correspond to a breathing vibration of an equilateral triangle as suggested in the ACM there should be additional negative parity parity states, such as $3^-$ and $4^-$, which so far have not been observed experimentally, even though "evidence has been found for $1^-$ and $3^-$ strengths associated with broad states between 11 and 14 MeV".  \cite{Freer2007}. 

Electric transition probabilities in the ground state band are given by \cite{acm}
\ba
B(EL;0^+ \rightarrow L^P) \;=\; \frac{(Ze \beta^L)^2}{4\pi} \frac{2L+1}{3} \left[ 1+2P_L(-\tfrac{1}{2}) \right] ~, 
\label{bel}
\ea
and the quadrupole moment of the first excited state by
\ba
Q_{2_1^+}  \;=\; \frac{2}{7} Ze \beta^2 ~.
\label{q2}
\ea
These expressions depend only on one coefficient, $\beta$, which is determined from the minimum in the elastic form factor to be $\beta=1.82$ fm. The ratio of 
quadrupole and octupole transitions depends on $\beta$
\ba
\frac{B(E3;3_{1}^{-} \rightarrow 0_{1}^{+})}{B(E2;2_{1}^{+} \rightarrow 0_{1}^{+})} 
\;=\; \frac{5}{2}  \beta^2 ~. 
\ea
Table~\ref{BELs} shows a comparison with the experimental data for $^{12}$C. The agreement is very good, especially if one takes into account that this calculation does not involve any additional parameter. The results are a direct consequence of the triangular configuration of $\alpha$-particles.

\begin{table}[t]
\caption{Electric transition probabilities and quadrupole moments in $^{12}$C and $^{13}$C.}
\label{BELs}
\centering
\begin{tabular}{lcrcl}
\hline
\noalign{\smallskip}
& $B(EL; i \rightarrow f)$ & Th & Exp & \\ 
\noalign{\smallskip}
\hline
\noalign{\smallskip}
$^{12}$C & $B(E2;2_{1}^{+}\rightarrow 0_{1}^{+})$ & $7.8$ & $7.63 \pm 0.19$ & $e^2\mbox{fm}^4$ \\ 
\noalign{\smallskip}
& $B(E3;3_{1}^{-}\rightarrow 0_{1}^{+})$ & $65.0$ & $104 \pm 14$ & $e^2\mbox{fm}^6$ \\ 
& $Q_{2_{1}^{+}}$ & $5.7$ & $5.97 \pm 0.30$ & $e\mbox{fm}^{2}$ \\
\noalign{\smallskip}
\hline
\noalign{\smallskip}
$^{13}$C & $B(E2;3/2_{1}^{-}\rightarrow 1/2_{1}^{-})$ & $7.8$ & $6.4 \pm 1.5$ & $e^2\mbox{fm}^4$ \\ 
\noalign{\smallskip}
& $B(E2;5/2_{1}^{-}\rightarrow 1/2_{1}^{-})$ & $7.8$ & $5.6 \pm 0.4$ & $e^2\mbox{fm}^4$ \\ 
\noalign{\smallskip}
& $B(E3;5/2_{1}^{+}\rightarrow 1/2_{1}^{-})$ & $65.0$ & $100 \pm 40$ & $e^2\mbox{fm}^6$ \\
& $Q_{5/2_{1}^{-}}$ & $5.7$ & & $e\mbox{fm}^{2}$ \\
& $Q_{3/2_{1}^{-}}$ & $4.0$ & & $e\mbox{fm}^{2}$ \\
\noalign{\smallskip}
\hline
\end{tabular}
\end{table}

\section{Cluster Shell Model} 

\begin{figure}
\centering
\begin{minipage}{.4\linewidth}
\centering
\includegraphics[width=0.75\linewidth]{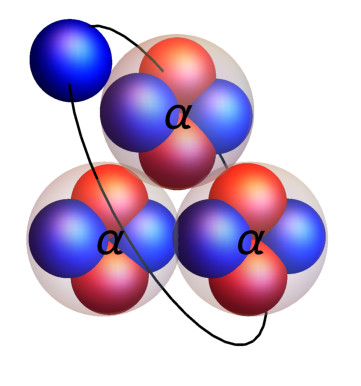}
\end{minipage}\hfill
\begin{minipage}{.55\linewidth}
\includegraphics[scale=0.6]{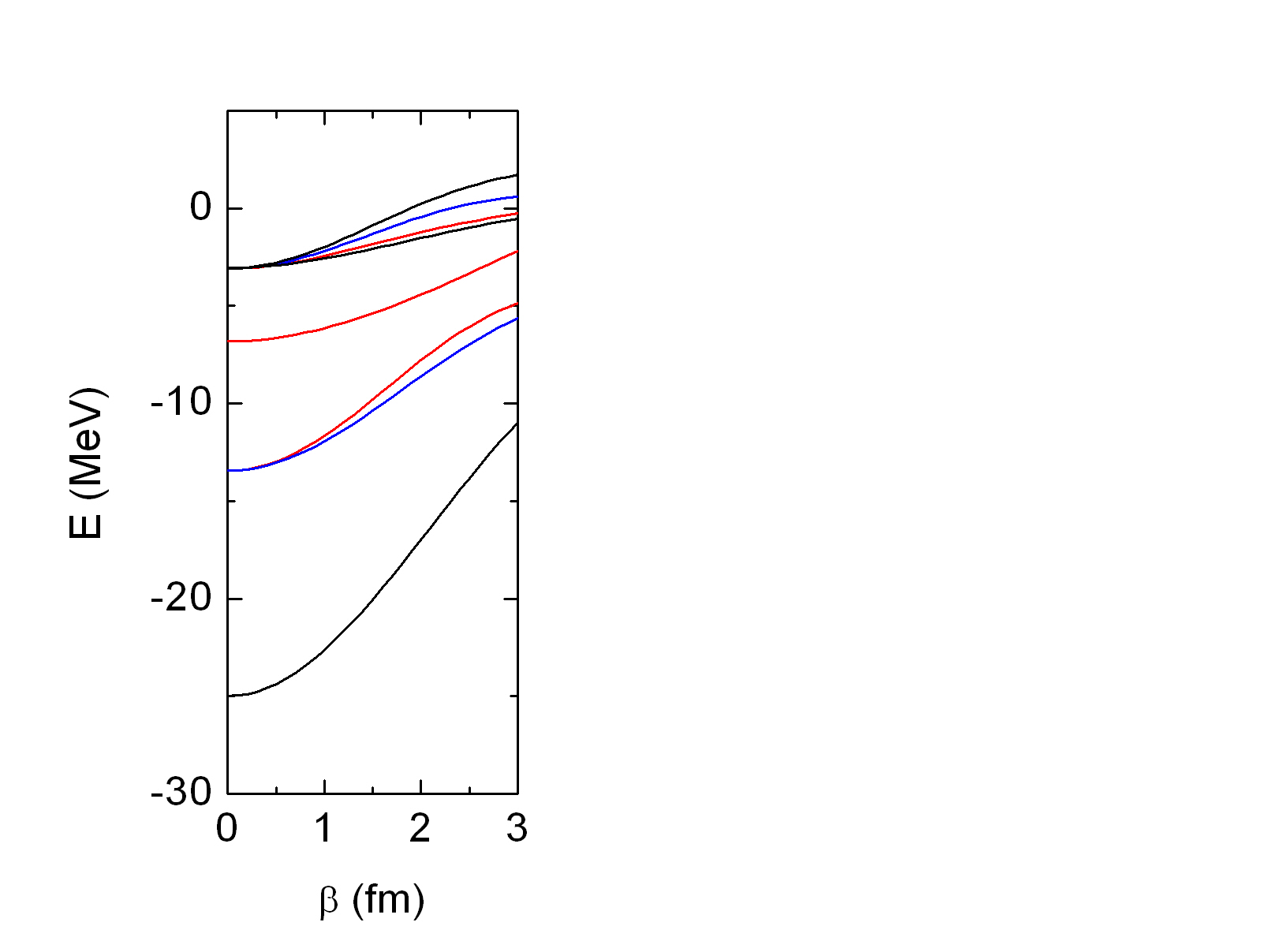}
\end{minipage}
\caption{Molecular-like picture of $^{13}$C. Taken from Ref.~\cite{prl122} (left).
Single-particle energies in a cluster potential with triangular symmetry (right).}
\label{splevels}
\end{figure}

The cluster shell model (CSM) was developed recently for the description of odd-cluster nuclei \cite{ppnp,csm}. The CSM combines cluster and single-particle degrees of freedom, and is very similar in spirit as the Nilsson model \cite{Nilsson}, but in the CSM the odd nucleon moves in the deformed field generated by the (collective) cluster degrees of freedom. The Hamiltonian is written as
\ba
H \;=\; T + V(\vec{r}) + V_{\rm so}(\vec{r}) + \tfrac{1}{2}(1+\tau_3) V_{\rm C}(\vec{r}) ~,
\ea
{\it i.e.} the sum of the kinetic energy, a central potential which is taken to be proportional to the $\alpha$-particle density, a spin-orbit interaction and, for an odd proton, a Coulomb potential. Fig.~\ref{splevels} shows the splitting of single-particles levels of a neutron moving in the deformed potential generated by a triangular configuration of three $\alpha$-particles as a function of $\beta$, the distance of the $\alpha$-particles from the center of mass $\beta$. The relevant value of $\beta$ is $1.82$ fm as determined from the minimum in the elastic form factor of $^{12}$C. The solutions of the CSM Hamiltonian are labeled by the irreducible representations of the double point group $D'_{3h}$: $\Omega=E_{1/2}$ (black), $E_{3/2}$ (blue) and $E_{5/2}$ (red), each of which is doubly degenerate. 

Figure~\ref{vibrations} shows the vibrational excitations for the coupling of a single-particle with $E_{5/2}$, $E_{1/2}$ and $E_{3/2}$ symmetry to the ground state band, the Hoyle band and the bending vibration in $^{12}$C. The levels marked with $\times$ have been identified in $^{12}$C and $^{13}$C. 

\begin{figure}
\centering
\setlength{\unitlength}{0.35pt}
\begin{picture}(720,430)(0,0)
\thicklines
\put (   0,0) {\line(1,0){720}}
\put (   0,400) {\line(1,0){720}}
\put (   0,0) {\line(0,1){400}}
\put (180,0) {\line(0,1){400}}
\put (360,0) {\line(0,1){400}}
\put (540,0) {\line(0,1){400}}
\put (720,0) {\line(0,1){400}}

\put ( 55,350) {\color{red} $\bf ^{12}C$}
\put (220,350) {\color{red} $\bf ^{12}C \otimes E_{5/2}$}
\put (400,350) {\color{red} $\bf ^{12}C \otimes E_{1/2}$}
\put (580,350) {\color{red} $\bf ^{12}C \otimes E_{3/2}$}

\put (100, 60) {\line(1,0){30}}
\put (100,200) {\line(1,0){30}}
\put (100,270) {\line(1,0){30}}
\put (110, 52) {$\times$}
\put (110,192) {$\times$}
\put (110,262) {$\times$}
\color{black}
\put (135, 55) {\color{blue} $\bf A'_1$}
\put (135,195) {\color{blue} $\bf A'_1$}
\put (135,265) {\color{blue} $\bf E'$}
\put ( 20, 57) {\bf Gsb}
\put ( 20,197) {\bf Hoyle}
\put ( 20,267) {\bf Bend}
\color{black}

\put (230, 60) {\line(1,0){30}}
\put (230,200) {\line(1,0){30}}
\put (230,265) {\line(1,0){30}}
\put (230,275) {\line(1,0){30}}
\put (240, 52) {$\times$}
\put (240,192) {$\times$}
\color{black}
\put (265, 55) {\color{blue} $\bf E_{5/2}$}
\put (265,195) {\color{blue} $\bf E_{5/2}$}
\put (265,250) {\color{blue} $\bf E_{1/2}$}
\put (265,275) {\color{blue} $\bf E_{3/2}$}
\color{black}

\put (410, 60) {\line(1,0){30}}
\put (410,200) {\line(1,0){30}}
\put (410,265) {\line(1,0){30}}
\put (410,275) {\line(1,0){30}}
\put (420, 52) {$\times$}
\color{black}
\put (445, 55) {\color{blue} $\bf E_{1/2}$}
\put (445,195) {\color{blue} $\bf E_{1/2}$}
\put (445,250) {\color{blue} $\bf E_{5/2}$}
\put (445,275) {\color{blue} $\bf E_{3/2}$}
\color{black}
\put (590, 60) {\line(1,0){30}}
\put (590,200) {\line(1,0){30}}
\put (590,265) {\line(1,0){30}}
\put (590,275) {\line(1,0){30}}
\color{black}
\put (625, 55) {\color{blue} $\bf E_{3/2}$}
\put (625,195) {\color{blue} $\bf E_{3/2}$}
\put (625,250) {\color{blue} $\bf E_{1/2}$}
\put (625,275) {\color{blue} $\bf E_{5/2}$}
\end{picture}
\caption{Vibrational excitations in nuclei with triangular symmetry.}
\label{vibrations}
\end{figure}

The structure of rotational bands for the case of a single-particle coupled to a triangular configuration of $\alpha$ particles is shown in Figure~\ref{rotbands}. The left panel shows the result for the ground-state band of even-cluster nuclei, as has been observed in $^{12}$C. The ground-state band consists of a series of $K$-bands with $K=3k$ ($k=0,1,2,\ldots,$) and angular momenta $L=0,2,4,\ldots,$ for $K=0$ and $L=K,K+1,K+2,\ldots,$ for $K \neq 0$. The parity is given by $P=(-1)^K$. The results for the case of the coupling of a single-particle level with $E_{5/2}$, $E_{1/2}$ and $E_{3/2}$ symmetry to the ground-state band of a triangular configuration with $A'_1$ symmetry is presented in the 2nd, 3rd and 4th panels of Fig.~\ref{rotbands}. Just as for even-cluster nuclei, each rotational sequence for odd-cluster nuclei consists of a series of $K$-bands given by 
\ba
E_{1/2} &:& K^P= 1/2^+, 5/2^-, 7/2^-, \ldots, 
\nonumber\\
E_{5/2} &:& K^P= 1/2^-, 5/2^+, 7/2^+, \ldots, 
\nonumber\\
E_{3/2} &:& K^P= 3/2^{\pm}, 9/2^{\pm}, \ldots,
\ea
with angular momenta $J=K,K+1,K+2,\ldots$.

\begin{figure}[b]
\setlength{\unitlength}{0.45pt}
\begin{picture}(790,430)(0,-30)
\thicklines
\put (  0,-30) {\line(1,0){790}}
\put (  0,400) {\line(1,0){790}}
\put (  0,-30) {\line(0,1){430}}
\put (160,-30) {\line(0,1){430}}
\put (390,-30) {\line(0,1){430}}
\put (620,-30) {\line(0,1){430}}
\put (790,-30) {\line(0,1){430}}

\put ( 60,350) {\color{red} $\bf ^{12}C$}
\put (230,350) {\color{red} $\bf ^{12}C \otimes E_{5/2}$}
\put (460,350) {\color{red} $\bf ^{12}C \otimes E_{1/2}$}
\put (660,350) {\color{red} $\bf ^{12}C \otimes E_{3/2}$}

\put ( 30, 60) {\line(1,0){20}}
\put ( 30,120) {\line(1,0){20}}
\put ( 30,260) {\line(1,0){20}}
\put ( 90,180) {\line(1,0){20}}
\put ( 90,260) {\line(1,0){20}}
\put ( 40, 10) {\color{blue} $\bf 0^+$}
\put (100, 10) {\color{blue} $\bf 3^-$}
\put ( 55, 55) {$\bf 0^+$}
\put ( 55,115) {$\bf 2^+$}
\put ( 55,255) {$\bf 4^+$}
\put (115,175) {$\bf 3^-$}
\put (115,255) {$\bf 4^-$}
\put (190, 60) {\line(1,0){20}}
\put (190, 90) {\line(1,0){20}}
\put (190,140) {\line(1,0){20}}
\put (190,210) {\line(1,0){20}}
\put (190,300) {\line(1,0){20}}
\put (250,140) {\line(1,0){20}}
\put (250,210) {\line(1,0){20}}
\put (250,300) {\line(1,0){20}}
\put (310,210) {\line(1,0){20}}
\put (310,300) {\line(1,0){20}}
\put (200, 10) {\color{blue} $\bf \frac{1}{2}^-$}
\put (215, 55) {$\bf \frac{1}{2}^-$}
\put (215, 85) {$\bf \frac{3}{2}^-$}
\put (215,135) {$\bf \frac{5}{2}^-$}
\put (215,205) {$\bf \frac{7}{2}^-$}
\put (215,295) {$\bf \frac{9}{2}^-$}
\put (260, 10) {\color{blue} $\bf \frac{5}{2}^+$}
\put (275,135) {$\bf \frac{5}{2}^+$}
\put (275,205) {$\bf \frac{7}{2}^+$}
\put (275,295) {$\bf \frac{9}{2}^+$}
\put (320, 10) {\color{blue} $\bf \frac{7}{2}^+$}
\put (335,205) {$\bf \frac{7}{2}^+$}
\put (335,295) {$\bf \frac{9}{2}^+$}

\put (420, 60) {\line(1,0){20}}
\put (420, 90) {\line(1,0){20}}
\put (420,140) {\line(1,0){20}}
\put (420,210) {\line(1,0){20}}
\put (420,300) {\line(1,0){20}}
\put (480,140) {\line(1,0){20}}
\put (480,210) {\line(1,0){20}}
\put (480,300) {\line(1,0){20}}
\put (540,210) {\line(1,0){20}}
\put (540,300) {\line(1,0){20}}
\put (430, 10) {\color{blue} $\bf \frac{1}{2}^+$}
\put (445, 55) {$\bf \frac{1}{2}^+$}
\put (445, 85) {$\bf \frac{3}{2}^+$}
\put (445,135) {$\bf \frac{5}{2}^+$}
\put (445,205) {$\bf \frac{7}{2}^+$}
\put (445,295) {$\bf \frac{9}{2}^+$}
\put (490, 10) {\color{blue} $\bf \frac{5}{2}^-$}
\put (505,135) {$\bf \frac{5}{2}^-$}
\put (505,205) {$\bf \frac{7}{2}^-$}
\put (505,295) {$\bf \frac{9}{2}^-$}
\put (550, 10) {\color{blue} $\bf \frac{7}{2}^-$}
\put (565,205) {$\bf \frac{7}{2}^-$}
\put (565,295) {$\bf \frac{9}{2}^-$}

\put (650, 90) {\line(1,0){20}}
\put (650,140) {\line(1,0){20}}
\put (650,210) {\line(1,0){20}}
\put (650,300) {\line(1,0){20}}
\put (710,300) {\line(1,0){20}}
\put (660, 10) {\color{blue} $\bf \frac{3}{2}^\pm$}
\put (675, 85) {$\bf \frac{3}{2}^\pm$}
\put (675,135) {$\bf \frac{5}{2}^\pm$}
\put (675,205) {$\bf \frac{7}{2}^\pm$}
\put (675,295) {$\bf \frac{9}{2}^\pm$}
\put (720, 10) {\color{blue} $\bf \frac{9}{2}^\pm$}
\put (735,295) {$\bf \frac{9}{2}^\pm$}
\end{picture}
\vspace{15pt}
\caption{Rotational bands for a triangular configuration of $\alpha$ particles 
in even-cluster nuclei (first panel) and odd-cluster nuclei with $E_{5/2}$, $E_{1/2}$ and 
$E_{3/2}$ symmetry (second, third and fourth panel). Each rotational band is labeled by the 
quantum numbers $K^P$.}
\label{rotbands}
\end{figure}

\begin{figure}
\centering
\includegraphics[width=0.75\linewidth]{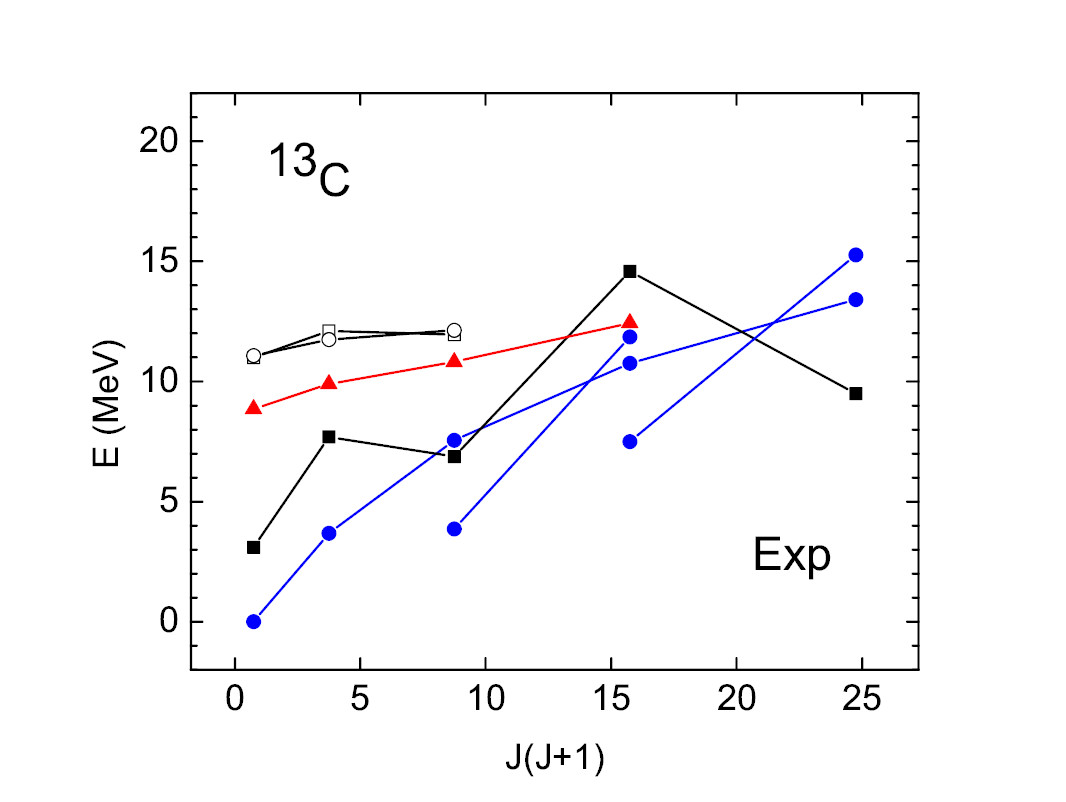}
\caption{Rotational bands in $^{13}$C. Taken from Ref.~\cite{prl122}.}
\label{bands_c13}
\end{figure}

In this case, the rotational spectrum is given by \cite{epjst}
\ba
E_{\Omega}(J) \;=\; \frac{1}{2{\cal I}} \left[ J(J+1) - 2K^2
+ \delta_{K,1/2} \, a_{\Omega} (-1)^{J+1/2} (J+\tfrac{1}{2}) \right] ~,
\label{erot}
\ea
where the last term denotes the Coriolis mixing. In Figure~\ref{bands_c13} we show the rotational band structure in $^{13}$C as suggested by our CSM analysis \cite{prl122}. The blue circles denote the members of the ground state rotational band with $E_{5/2}$ symmetry and $K^P=1/2^-$, $5/2^+$ and $7/2^+$. The $J^P=1/2^+$ state at $3.089$ MeV is assigned as the bandhead of the rotational band with $E_{1/2}$ symmetry and $K^P=1/2^+$ (black squares), and the $J^P=1/2^-$ state at $8.860$ MeV as the bandhead of the analogue of the Hoyle band (red triangles). 

In the CSM, the transition probabilities consist of a contribution from the collective (cluster) part and the single-particle part. For electric transitions, the transition probabilities are dominated by the collective part. If we neglect the single-particle part, we find the following correlations between the even- and odd-cluster nuclei 
\ba
B(E2;3/2_{1}^{-} \rightarrow 1/2_{1}^{-})  &=&
B(E2;5/2_{1}^{-} \rightarrow 1/2_{1}^{-}) \;=\;  B(E2;2_{1}^{+}\rightarrow 0_{1}^{+}) ~,
\nonumber\\ 
B(E3;5/2_{1}^{+} \rightarrow 1/2_{1}^{-}) &=& B(E3;3_{1}^{-}\rightarrow 0_{1}^{+}) ~,
\nonumber\\
Q_{5/2_1^-} &=& \tfrac{10}{7} \, Q_{3/2_1^-} \;=\; Q_{2_1^+} ~.
\label{correlations}
\ea
Table~\ref{BELs} shows that to a good approximation the experimental data in $^{12}$C and $^{13}$C can be understood by Eqs.~(\ref{bel},\ref{q2},\ref{correlations}).

\begin{figure}[t]
\centering
\includegraphics[width=0.6\linewidth]{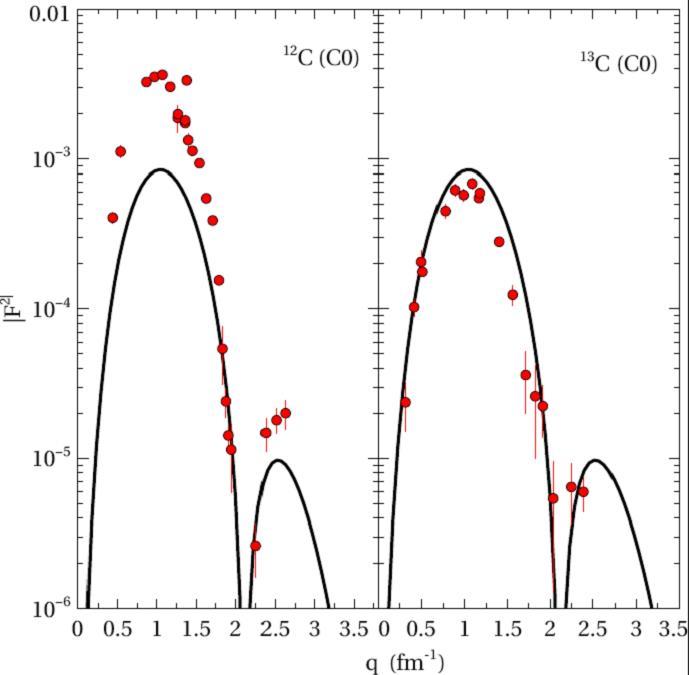}
\caption{Form factor to the $0_2^+$ Hoyle state in $^{12}$C (left) and the $\frac{1}{2}_2^-$ state at 8.86 MeV in $^{13}$C (right). Taken from Ref.~\cite{plb843}.}
\label{hoyle_c13}
\end{figure}

Finally, the $C0$ form factors of the Hoyle state and its analogue in $^{13}$C are expected to be the same \cite{plb843}. Figure~\ref{hoyle_c13} confirms the interpretation of the $J^P=1/2^-$ state at $8.860$ MeV  in $^{13}$C as the analogue of the Hoyle state.

\section{Summary and conclusions}

We presented a review of the Algebraic Cluster Model and the Cluster Shell Model. The two models were applied to the cluster states in $^{12}$C and $^{13}$C, respectively. An analysis of both the rotation-vibration spectra and electromagnetic transition rates shows strong evidence for a cluster structure of three $\alpha$-particles in a triangular configuration. In both cases, the ground state band consist of positive and negative parity states which merge to form a single rotational band. This interpretation is validated by the observance of strong $B(EL)$ values within the ground state band. The rotational sequences can be considered as the fingerprints of the underlying geometric configuration (or point-group symmetry) of $\alpha$ particles. 

The CSM was introduced to describe the properties of cluster nuclei with $A=4k+x$ in terms of a cluster of $k$ $\alpha$-particles plus $x$ extra nucleons. For $k=0$ it reduces to the ACM. In this contribution, we focussed on $\alpha$-cluster nuclei with $A=4k$ nucleons, such as $^{12}$C  and odd-cluster nuclei, such as $^{13}$C with $x=1$. Currently, we are studying $\alpha$-cluster nuclei with $x=2$ extra nucleons to investigate the interesting question to what  extent  the cluster structure of $\alpha$-particles persists under the addition of two neutrons. As a first example, we studied the nucleus $^{10}$Be as a cluster of $k=2$ $\alpha$-particles and $x=2$ neutrons \cite{be10}. 

\section*{Acknowledgments}

This work was supported in part by PAPIIT-DGAPA grant IG101423 (Mexico).

\end{document}